\documentclass[conference, compsocconf]{IEEEtran}

\usepackage{listing}
\usepackage{eepic}
\usepackage{epic}
\usepackage{xspace}    
\usepackage{graphicx}
\usepackage{latexsym}
\usepackage[footnotesize]{subfigure}
\usepackage{afterpage}
\usepackage{amstext}   
\usepackage{url}
\usepackage{alltt}
\usepackage{booktabs}
\usepackage{epstopdf}
\usepackage{xcolor}
\usepackage{color}
\usepackage{morefloats}
\usepackage{footmisc}
\usepackage{amsmath}
\usepackage{multirow}
\usepackage{pifont} 
\usepackage{amsmath}
\usepackage{multicol}
\usepackage[figuresright]{rotating}
\usepackage{bm}
\usepackage{caption}
\usepackage{fancyhdr}
\usepackage{geometry}
\usepackage{amsmath}
\usepackage{amsfonts}
\usepackage{setspace}
\usepackage{array}
\usepackage{amssymb}
\usepackage{graphicx}
\usepackage{url}
\usepackage{multicol}
\usepackage{stfloats}
\usepackage{subfigure}
\usepackage{changes}
\usepackage{eurosym}
\usepackage{rotating}
\usepackage{color}
\usepackage{graphicx}
\newcommand{\bi}{\begin{itemize}}
	\newcommand{\ei}{\end{itemize}}
\newcommand {\beq}{\begin{equation}}
\newcommand {\eeq}{\end{equation}}
\newcommand {\be}{\begin{enumerate}}
	\newcommand {\ee}{\end{enumerate}}

\begin{document}
\title{Middleware Technologies for Cloud of Things - a survey}

\author{\IEEEauthorblockN{
		Amirhossein Farahzadi\IEEEauthorrefmark{1},
		Pooyan Shams\IEEEauthorrefmark{1},
		Javad Rezazadeh\IEEEauthorrefmark{1}\IEEEauthorrefmark{4},
		Reza Farahbakhsh\IEEEauthorrefmark{3}}
	\IEEEauthorblockA{\IEEEauthorrefmark{4}University of Technology Sydney, Australia.
	\{rezazadeh@ieee.org\}}
	\IEEEauthorblockA{\IEEEauthorrefmark{3}Institut Mines-Telecom, Telecom Sud-Paris, CNRS UMR 5157, France.\\
		\{reza.farahbakhsh, noel.crespi\}@it-sudparis.eu} 
	\IEEEauthorblockA{\IEEEauthorrefmark{1}Islamic Azad University, North Tehran Branch, Tehran, Iran\\
		E-mail: \{a.farahzadi@iau-tnb.ac.ir, pooyan.shams@gmail.com}\\
}

%
%
%

\maketitle
\begin{abstract}
The next wave of communication and applications rely on the new services provided by Internet of Things which is becoming an important aspect in human and machines future. The IoT services are a key solution for providing smart environments in homes, buildings and cities. In the era of a massive number of connected things and objects with a high grow rate, several challenges have been raised such as management, aggregation and storage for big produced data. In order to tackle some of these issues, cloud computing emerged to IoT as Cloud of Things (CoT) which provides virtually unlimited cloud services to enhance the large scale IoT platforms.
There are several factors to be considered in design and implementation of a CoT platform. One of the most important and challenging problems is the heterogeneity of different objects.
This problem can be addressed by deploying suitable "Middleware". Indeed, Middleware sits between things and applications that make a reliable platform for communication among things with different interfaces, operating systems, and architectures.
The main aim of this paper is to study the middleware technologies for CoT. Toward this end, we first present the main features and characteristics of middlewares. Next we study different architecture styles and service domains. Then we presents several middlewares that are suitable for CoT based platforms and lastly a list of current challenges and issues in design of CoT based middlewares is discussed.
\end{abstract}





\section{Introduction}
\label{sec:Introduction}
The appearance of Internet of Things (IoT) concept is shaping and reshaping how future services are going to be define. The main idea behind this concept is to develop different type of communication network based on group of physical objects or simply ``things". The IoT objects embedded with electronic chips, software, sensors and internet connectivity to collect and process data from the environment or affecting it by deploying actuators. IoT combines real-world data and computer processing to lower the costs and increase the efficiency and accuracy. Each thing can be recognized separately through its embedded computing system and is able to communicate with other things through Internet infrastructure.
Recently the number of connected and embedded smart devices grows rapidly.
According to Cisco IBSG \cite{Cisco}, IoT world will includes more than 50 billion objects in 2020.

IoT is translated in different concepts or approach such as ``Network-Oriented" or ``Object-Oriented" or even, as it mentioned in \cite{Atzori:2010:ITS:1862461.1862541}, ``Semantic Oriented". These visions emerged because of different stakeholder ideas because different vendors and IT experts have their own vision of this technology. IoT semantically means ``a worldwide network of interconnected objects uniquely addressable based on standard communication protocols" \cite{roadmap}. International Telecommunication Union (ITU) also defines IoT as a network which provides connectivity ``anytime, anyplace for any connected smart devices".

\begin{figure}[t]
	\centering
	\includegraphics[width=0.4\textwidth,height=0.4\textwidth]{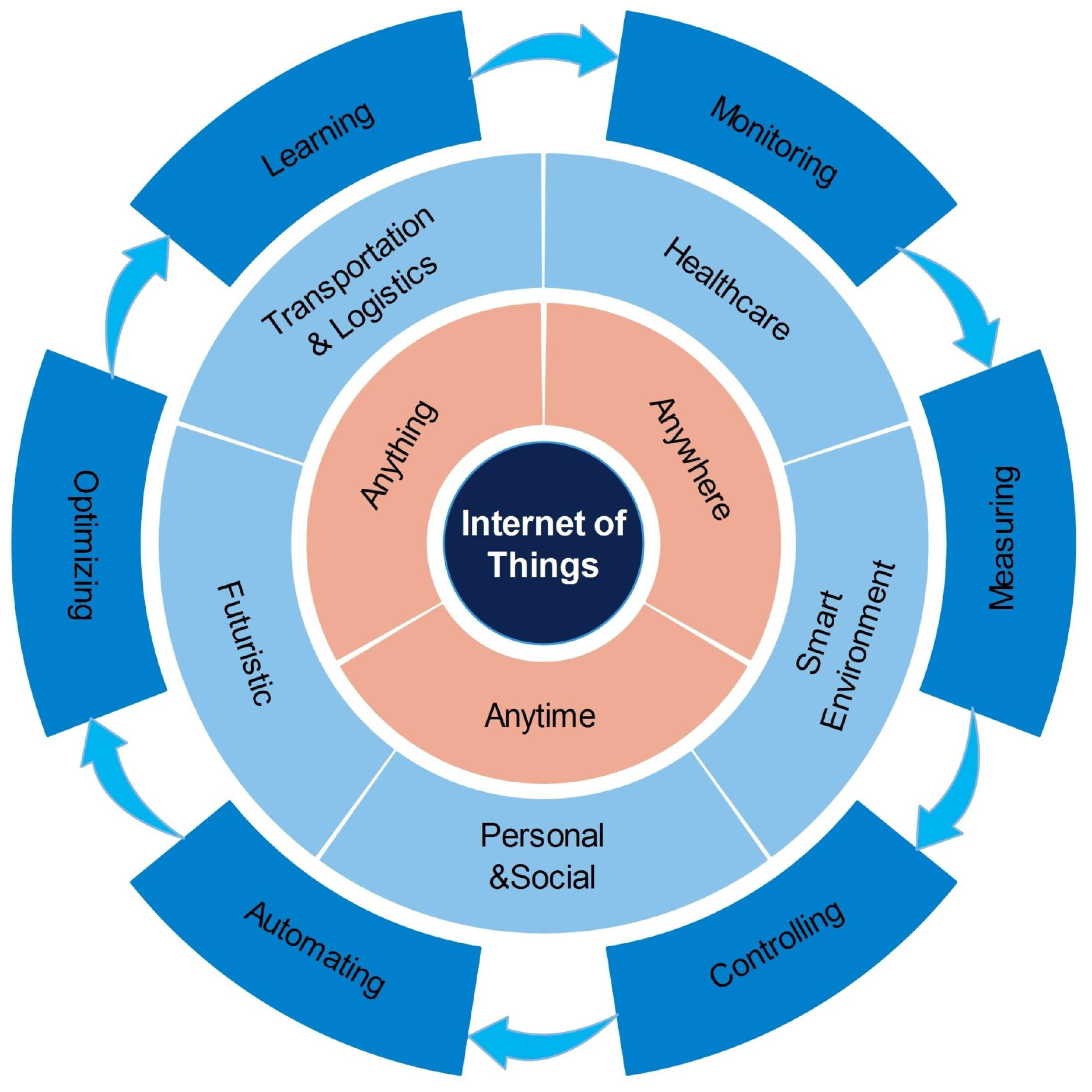}
	\caption{Internet of Things - main concept and functionalities}
	\label{fig1:IoT}
	\vspace{-0.3cm}
\end{figure}

Figure \ref{fig1:IoT} shows a high level concept of IoT \cite{knuthwebsite2} including the main concept and its high level functionalities.
As shown, the IoT main characteristics is presented in the core circle of the figure including anywhere, anything and anytime features that indicate limitless IoT realm. It is noteworthy to know applying this technology with CPS (Cyber Physical Systems) and Cloud Computing creates Industry 4.0. 
The middle circle includes the general application domains. Cross-system automation makes tasks to be performed more accurately, coordinately, and conveniently. In the outer circle, we illustrate a general cyber-physical learning process. Each cycle of this process boost systems knowledge and performance.
First, monitoring operations carried out by sensors. Then the system will measure and store the data. In the control phase, it will check whether measured data has touched or passed pre-defined minimums, maximums or thresholds. Next, according to controlled data, the system decides which automated tasks should be performed. These tasks can be a typical operation, error handling, alerting, and etc. Optimization tries to fix the problems, defects or simply makes the system to perform better. The last section is learning phase which helps to improve system knowledge and documentation.

\begin{figure}[t]
	\centering
	\includegraphics[width=0.4\textwidth,height=0.35\textwidth]{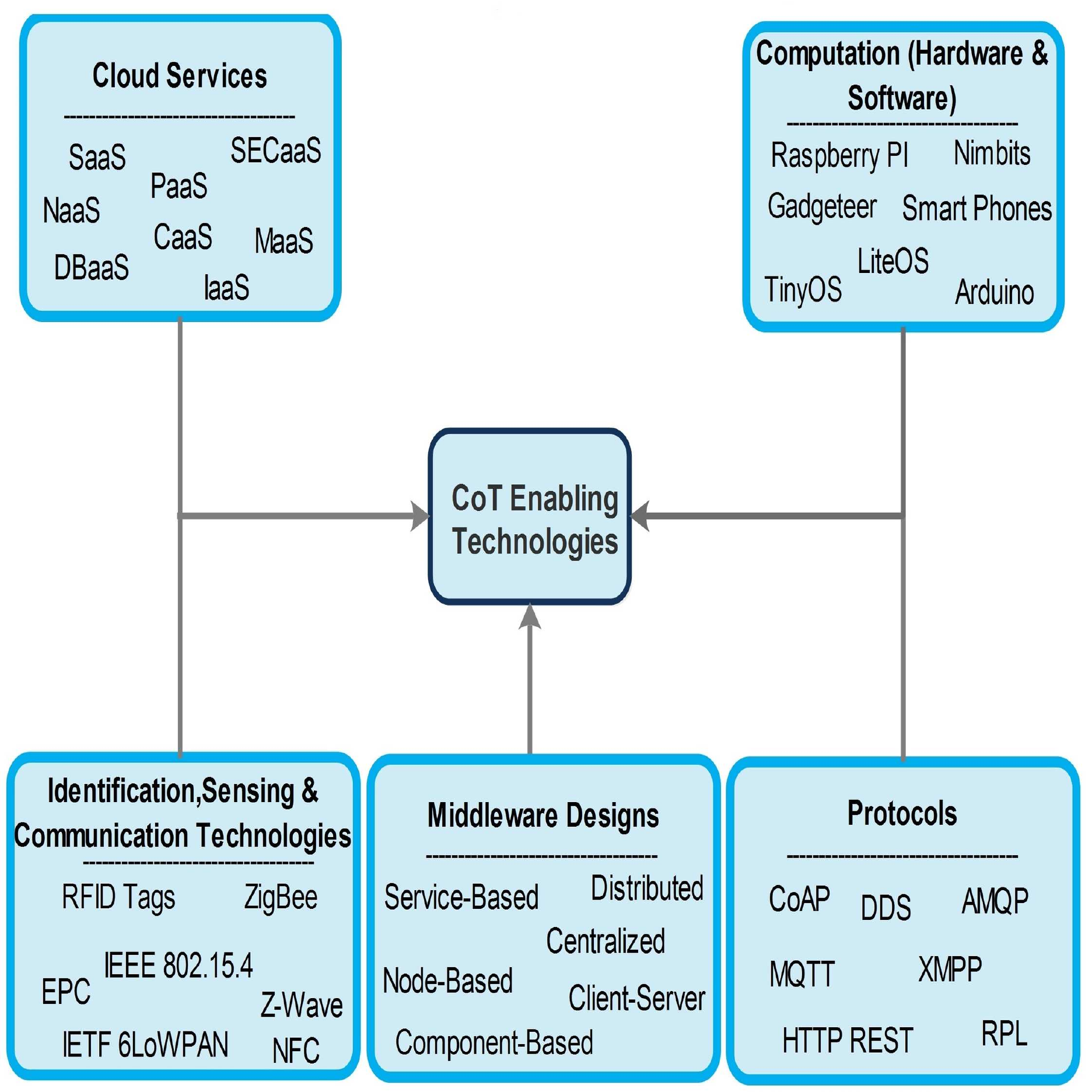}
	\caption{CoT enabling technologies}
	\label{fig2:Enable}
	\vspace{-0.3cm}
\end{figure}

Considering this huge population and the fact that all connected objects actively produce or request data and require various services, the issue of limited available resources is a key factor to be considered in design of large scale IoT infrastructure In addition, other issues such as scalability, storage capacity and maintainability are significant challenges as well.
In this era, cloud computing became an efficient, accessible, and reasonable solution. IoT in combination with Cloud functionalities provide a new phenomena called as \emph{Cloud of Things (CoT)} where enable many new possibilities such as Big Data processing as well as covering security concerns, resource constraints, and scalability to some extend. CoT creates new revenue streams, improve customer services and inspire product innovations. As Figure \ref{fig2:Enable} shows, CoT can be considered as a composition of at least this five enabling technologies. Among these enablers, middleware play a key role in CoT (similarly to IoT). 

The main focus of this paper is on middleware technologies and it aims to study their roles and necessities in CoT environment.
Middleware is a software layer that sits between applications and the objects. It aims to provide solutions to frequently encountered problems such as heterogeneity, interoperability, security and dependability \cite{Issarny:2007:PFM:1253532.1254722}?.
We can consider middleware in ``network-oriented" vision according to \cite{Gubbi:2013:ITV:2489313.2489456}.

Every day we are witnessing growth in middlewares development because this enabler makes it easier to combine new services and previous technologies to produce a novel and more capable one.
Transparency is the main feature that middlewares can offer. It provides an abstraction to the applications from different objects and this feature will solve architecture mismatch problems. There are some other features that middlewares should present for the desirable performance, flexibility, context management, interoperability, reusability, portability, maintainability and a few more properties that we will explain them in this paper.

This paper is organized as follow. Next section (Sec. \ref{sec:related_work}) provides a brief overview of the related studies. Section \ref{features} discuss features and characteristics of middlewares followed by Section \ref{arch_comparison} which describes different architecture designs for middlewares.
Section \ref{service_domain} is devoted to middleware service domains and their applications. In section \ref{middlewares}, we review a number of middlewares and compared them based on different features. Lastly, we discuss few challenges and issues in Section \ref{challenges} and conclude this paper in Section \ref{Conclusion}.

\section{Literature Review}
\label{sec:related_work}
A large and growing body of literature has investigated IoT and Cloud Computing. Most of these studies focused on the combination of aforesaid domains to bring up a novel and mature technology \cite{6581037}\cite{Botta2016684}\cite{6461705}\cite{6742575}.
As an example, Mohammad Aazam et al introduced CoT in \cite{6778179} and explained its necessity. In addition several key issues on integration of IoT and Cloud Computing are discussed including data management, security and privacy, resource allocation, identity management, etc. In another work \cite{noto2016clouds}, authors show moving toward CoT is essential and can help to implement smart environments more efficiently. The main concerns here are data protection, privacy, and consumer law. CoT can utilize system performance by taking advantage of cloud services but, exchanging massive control or data packets can be harmful to this system and make it less efficient. There are situations which exchanging data between IoT and Cloud is not reasonable (i.e. requesting simple services or storing temporary data in the cloud). Therefore, here \cite{6827673} it will present a smart gateway to process and analysis requests and decides whether to answer them locally or sends them to the cloud.

There is a relatively small body of literature that is concerned with middlewares in CoT or even in IoT. Authors in \cite{Gama:2012:CHS:2125388.2125412} discuss what benefits can be offered by Service Oriented Computing (SOC) to build a middleware for the Internet of Things. Some concepts have been applied in a Service Oriented Architecture (SOA) middleware that tries to leverage the existing IoT architectural concepts by using SOC features in order to provide more flexibility and dynamicity.

In \cite{5678448}, authors present a new application layer resolution for interoperability. The key concept is to utilize device semantics provided by available specifications and dynamically wrap them into middleware as semantic services. In this paper \cite{perera2013context}, it presents CASSARAM, a Context-aware sensor probe, Selection and Ranking model for IoT to address the research challenges of choosing sensors when large numbers of sensors with overlapping and sometimes redundant In \cite{6758734}, authors propose Mobile Sensor Data Processing Engine (MOSDEN), a plug-in-based IoT middleware for mobile devices that tries to collect and process sensor data without programming endeavors. This architecture also supports sensing as a service model. Another paper \cite{6296229} has worked on e-health care domain. E-health features include tracking, identification, authentication, data collection and sensing. VIRTUS is an Instant Messaging Protocol (XMPP)-based middleware which tries to provide a real-time, safe and trustworthy communication channel among heterogeneous devices. CASP is another Context gathering framework which considers some essential requirements in order to act properly. A programming interface, active/passive sensor mode, simplicity, multi-transport support, separation of concerns and sensor data model are some vital requirements in this framework ? \cite{Devaraju:2007:CGF:1378063.1378070}.

Apart from the mentioned academic studies, several commercial CoT platforms have been developed including GroveStreams\footnote{https://grovestreams.com/}, EVRYTHNG\footnote{https://evrythng.com/}, and Fusion Connect\footnote{http://www.fusionconnect.com/} projects. GroveStreams is a platform which can process Big Data from a wide range of devices. It can provide data analytics tools nearly in real-time. GroveStreams supports different industrial domains and by applying cloud services, it can convert received raw data into meaningful information. EVRYTHNG platform is suitable for digital identification of products so they can exchange data and information with authorized management applications. This platform guarantees SLA by using end-to-end secure, reliable and flexible management. Fusion Connect is a remarkable no-coding CoT platform. Every operation happens by the drag-and-drop approach. By deploying this platform, users are able to virtualize things, connect them to reporting devices and perform analytics to unlock their data. It can provide multiple services such as predicting products failure, automation of maintenance operations, producing performance statistics related to objects, optimizing supply chain and calculate material replenishment costs.

In view of all that has been mentioned so far, there in no dedicated study on middleware in CoT which is the main aim of this study.

\section{Features and Characteristics of Middlewares}
\label{features}
\begin{figure}[t]
	\centering
	\includegraphics[width=0.4\textwidth, height=0.53\textwidth]{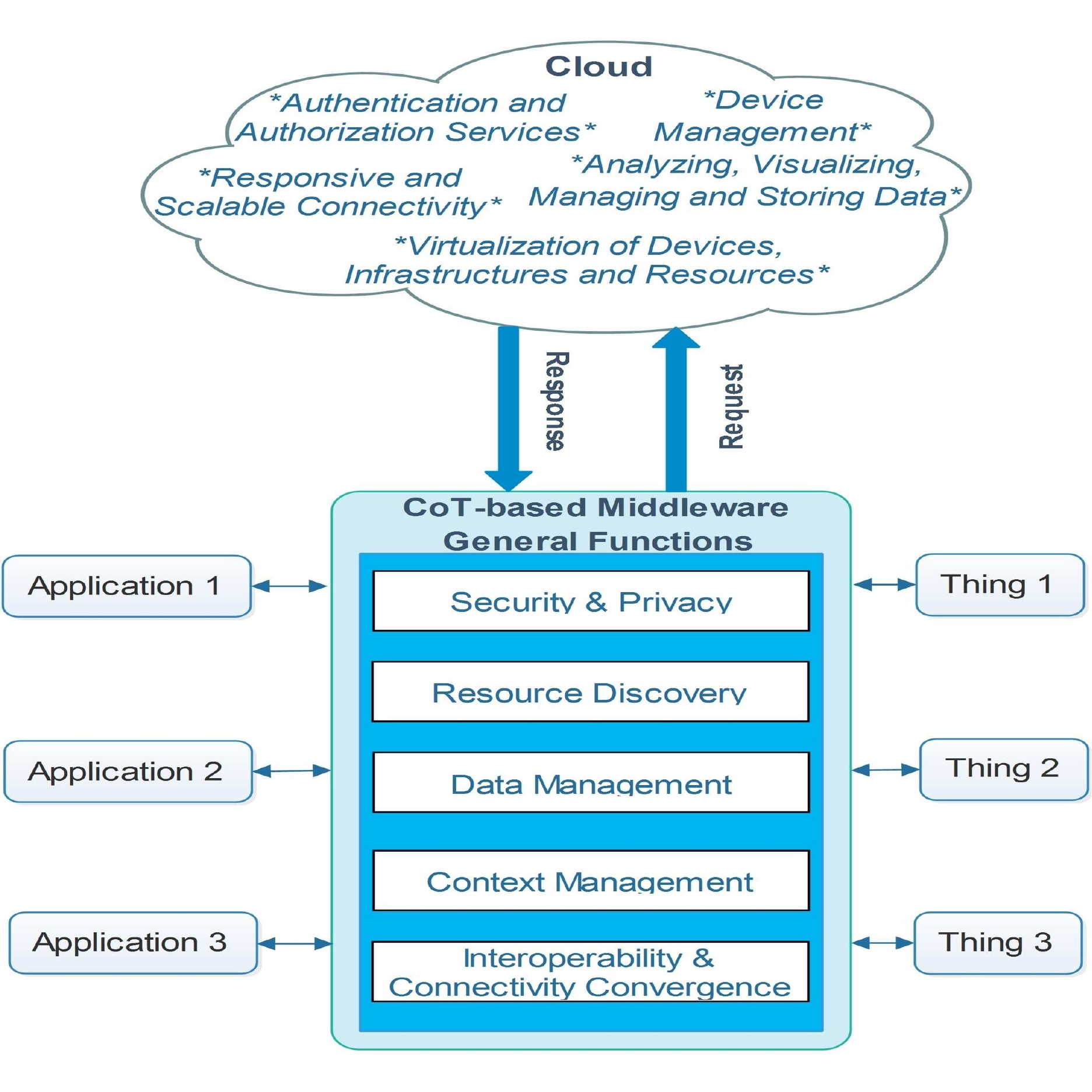}
	\vspace{-0.2cm}
	\caption{CoT-based Middleware - overall concept, its position and the main functionalities}
	\label{fig3:middleware}
	\vspace{-0.3cm}
\end{figure}

As mentioned earlier, middleware is an important enabler which provide communication among heterogeneous things. It is a mid-layer between things and application services and provides an abstraction of the thing's functionality for application services. Figure \ref{fig3:middleware} shows a general vision of CoT-based middleware and includes the overall concept of it as well as its position in the design and the main functionalities that will be provided by a middleware.

As shown, a middleware is able to bring flexibility and several features and characteristics that we can see various combinations of them according to the system's requirements. In the following parts, we discuss number of features and characteristics of middlewares.



\subsection{Flexibility}
Flexibility is one of the most important capabilities that a middleware can offer to IoT or CoT systems. By using middleware, part of application developers' concerns are covered because it can handle conflicting issues due to communication between applications and things. 
There are different kinds of flexibility (e.g., response time or delay flexibility). As a result, flexibility moves from application level to middleware level in order to handle different forms of flexibility.
It is essential to determine which software or hardware components needs more flexibility. The level of flexibility is important because high level of flexibility means more connectivity APIs and processing workforce which is not reasonable for multiple domains such as ultra resource constrained IoT networks.

\subsection{Transparency}
Middleware hides many complexities and architectural information details from both application and object sides, so they can communicate with the minimum knowledge of other side's necessary details of information.Transparency is a distinctive feature of middleware that can be very helpful for programmers, end users and applications. According to the application domain and services, developers need to decide which aforementioned part requires transparency the most.
Middleware transparency is in forms of platform and network.
\emph{(i) Platform transparency}: middleware runs on a variety of platforms, it lets organization use different hardware platforms according to their requirements. Clients and servers do not need previous knowledge to work with each other.
\emph{(ii) Network transparency}: middleware provide transparency of the networks to the application users. This means users don? need to know whether resources are located locally or on remote devices.

\subsection{Context Management}
This feature is a coordination management concept. It allows users to choose and configure a service or subject in one application, and then all other applications which containing information about that specific subject will adjust themselves with the same setting that user defined previously. This feature is the main pillar in context-aware systems. In order to achieve high level of flexibility which is important for better overall performance, context management allows numerous programs' thread to work on same task, a thread to work on different task or each thread to perform different tasks.
CoT is a set of ubiquitous devices, it is necessary to know the diversity of objects technically, logically and physically. This will lead to proper context management.

\subsection{Interoperability}
This functionality means two set of applications, on interconnected networks, be able to exchange data and services meaningfully with a different assumption about protocols, data models, configuration and etc, without any problems and additional programming efforts by developers. In fact the nature of software and hardware heterogeneity of CoT requires of having interoperable components. Interoperability will contribute to standardization. Interoperability has different degrees which depend on heterogeneity level of the environment or system components.
Implementing interoperability requires creation, management, reception and fulfillment of realistic standards. It is important to make sure that they are SMART.

\subsection{Re-usability}
On SOA based middlewares, there is the possibility to reuse software and hardware because there is no specific technology to impose its policies for service implementation \cite{Atzori:2010:ITS:1862461.1862541}. The main purpose of reusability is to make designing and developments easier by modifying system components and assets for specific requirements, which brings forth cost efficiency. Cloud-based networks are really crucial for obtaining this feature. These types of systems offer everything's as a service which can be used multiple times.

\subsection{Platform Portability}
Portability is a critical feature in CoT. There are many mobile hardware and software components which are constantly moving between different platforms. Therefore a CoT platform should be able to communicate from everywhere, anytime with any device. Therefore middleware helps to boost portability by its flexible manner. It helps to move across the environment and not being restricted to one platform.Actually middleware runs in user side and can provide independence from network protocol, programming language, OS, etc.
By considering platform variations in CoT, it is obvious that the applications and services need a kind of self-determining mechanism for cross-platform development. This can be achieved by using platform independence or portability. The difference is the first one run the applications on different platforms by using virtual machines to execute the codes and the later one is expected to adapt the applications to the new environment with reasonable users efforts.

\subsection{Maintainability}
This is an ability of systems, applications or devices to respond to failures properly and rapidly return to normal functionality without any problem. This can be achieved by isolating, correcting, repairing, preventing and other acts of resolving defects. Maintainability has a fault tolerance approximation. For performing maintainability efficiently it is necessary to have well-defined procedures and infrastructures. This feature is really needed in hard-real time applications.
The growth of deploying CoT in different contexts is significant. Without proper maintainability in middleware, extending the network or fixing the bugs would be overwhelming. Designing a maintainable middleware from outset is important. Providing relevant documentations and feedback checklists can be helpful for designing stable and extendable middleware.

\subsection{Resource Discovery}
An CoT system includes multiple heterogeneous devices.There is no reliable and global knowledge about availability of mentioned devices. Resource discovery is actually a process that used by a node to search and probe for intended resources such as services or data types among entire nodes of a network. After sending a search query, resource discovery protocol will automatically choose the best resource that offers the most effective services and information \cite{Atzori:2010:ITS:1862461.1862541}. For example, the process of choosing the best sensor in the environment that gives us the most reliable information.
One of the key essentials of a Resource Discovery mechanism for middlewares is to cope with frequent failures. Developers must use self-stabilizing algorithms in order to bring the system into an ordinary state despite transient failures.

\subsection{Trustworthiness Management}
This capability is required in social Internet of Things applications. By this feature, we can develop trust and credit mechanisms (e.g., authentication, hashing, encryption and etc) to ensure that services and information that have been prepared for us by other applications, came from trustworthy systems \cite{6714496}. It plays a key role in establishing trust relationships among different applications. This will lead to a trustable, robust and secure system.

\subsection{Adaptability}
In CoT systems, environment and networks infrastructure often changes due to different reasons including nodes mobility, power drain, topology shifts and etc. Many middlewares usually has fixed capabilities and cannot be customized to dynamic and unpredictable situations. This feature shows how middleware should behave against environmental changes. It can react both statically and dynamically. Deploying optimized dynamic methods result better adaptability. Higher adaptability is critical in hard real-time systems. In contrary with flexibility, being adaptable means durability against long-term changes in the systems.
For reaching a reasonable amount of adaptability, it is necessary to provide it as services (such as security and identification protocols adaptation services, communication protocols adaptation services, semantic protocols adaptation services, etc).

\begin{table*}[t]
	\centering
	\caption{Comparison of different possible architecture designs in middlewares}
	\scriptsize
	\vspace{-0.2cm}
	\begin{tabular}{|l|m{6.4cm}|m{6.4cm}|}
		\hline
		\textbf{Architecture} & \textbf{Benefits} & \textbf{Challenges} \\
		\hline
		Component-based & Reusability, Abstraction support and Independency & Maintenance, Migration, Complexity and Compatibility \\
		\hline
		Distributed & Resource sharing, Openness, Scalability, Concurrency, Consistency and Fault tolerance & Interoperability, Security, Manageability and Maintainability \\
		\hline
		Service-based & Reusability, Scalability, Availability and Platform independence & Service discovery, Complex service management and Service identification \\
		\hline
		Node-based & Availability and Mobility & Security and Manageability \\
		\hline
		Centralized & Simplicity, Security and Manageability & Scalability, Availability and Portability \\
		\hline
		Client-server & Servers separation, Resource accessibility, Security, Back-up and Recovery & Congestion, Limited scalability and Single Point of failure \\
		\hline
	\end{tabular}%
	\label{tab:1}
\end{table*}

\subsection{Security and Privacy}
Security in middleware is a vital issue because most data transmission and operation are occurring through it. For having a secure system we must consider confidentially, integrity and availability.  Therefore, middleware should provide different security measures for ubiquitous applications and pervasive environment such as authentication where identification and credentials, authorized modifications and access control policy are needed for verification and managements in accountability \cite{AlJaroodi2010187}. Privacy means all IoT system components that access personal user information must guarantee protection of mentioned information from unauthorized access.

\subsection{Connectivity Convergence}
Internet of Things contains and supports various types of hardware and software components that interact with each other through heterogeneous communication platforms. For example, software applications may send their request or queries to middleware in different kinds of communication methods such as WiFi Signals, Wired-based data, Fiber-optic lights, Bluetooth and etc. On another side of the scheme, objects may use communication technologies like Zigbee and RFID. Even it is possible that sensors send directly analog signals or send digital signals after processing them. So it is obvious connectivity APIs and management is an essential need. This feature is convergent with interoperability.

\section{Architectural Comparison of Middlewares}
\label{arch_comparison}
Middlewares are designed based on different architectures. One of the most important aspects in design of a middleware is determining a well-defined framework or architecture. Each architecture has unique attributes therefore, they can be categorized based on different parameters as oppose to make an explicit and fixed categorization. In \cite{6512846}, C. Perera et al, proposed a general category for middle-ware's architectures as follow: Distributed architecture, Component-based architecture, Service-based architecture, Node-based architecture, Centralized architecture and Client-server architecture. The main features of the mentioned architectures are summarized in Table \ref{tab:1}. This table includes benefits and challenges for each architecture design in middlewares.
To provide more detail, Table \ref{tab:2} categorized a list of the well-known middlewares based on their application domains and architecture styles. The detail of each mentioned middleware is provided in Section \ref{middlewares}.


\subsection{Component-based}
First, we explain Component-based architecture. In this type of architecture, there are some specific and main loosely couple independent components that semantically are working together to perform tasks but technically, each one of them is responsible for solving a specific part of the problem. So it is often said that components are cohesive and modular. It decomposes the design based on logical or functional components which provide a higher level of abstraction. The principal objective of this architecture is the encapsulation of functionality and behaviors of system elements for minimal dependency, more reusability, and easier trouble-shooting.

\subsection{Distributed}
Distributed architectures or distributed systems consist of different networked software and hard-ware components that coordinate their operations to execute tasks. The main characteristics of this architecture are concurrency of components, lack of global clock and the fact that failure of a component doesn't affect the whole system. The main advantages are fault tolerance, scalability, concurrency, flexibility and hardware and software sharing.
Components with different platforms and on several machines can cooperate on the specific goal. There are several frameworks for supporting this architecture such as CORBA, .NET, J2EE, and etc.

\subsection{Service-based}
This architecture is one of the most efficient designing styles. There are two methods to implement SOA-based middleware. First is deploying it in stand-alone manner and second is using Cloud Computing services (PaaS).The components of this architecture provide services to each other over a communication protocol \cite{5175920}. Every device offers its functionality as standard services, while the detection and invocation of new functionalities from other services could be performed simultaneously. This architecture is not recommended for following applications
Homogeneous system: This architecture is not cost effective and practical for a single vendor.
Real-Time: As SOA requires synchronous communication between the service consumer and producer; it is not a suitable option for devices which need strictly-enforced response time such as embedded equipment. In this case, tighten coupled architectures is preferred.
GUI-based: SOA would not be a desirable option for GUI functional applications like maps which require heavy data traffic exchange.

\subsection{Node-based}
In this architecture, there are many software components with same or different functionalities that work on mobile and sensor networks in order to communicate and process data which collected from the sensors? \cite{6512846}. This architecture is composed of streams and nodes. Nodes operate the data through streams with other nodes and it makes them suitable for mobile devices.

\subsection{Centralized}
In this architecture all the services have gathered in a specific location and applications or devices can make a request to use that resources and services. Users are usually simple thin devices that hand over their request to the central server which is a resource-rich device. The network between devices and central server can be implemented by connectionless or connection oriented protocols although there is no direct communication between application instances. This architecture is the exact opposite of distributed architecture. The failure of central server will cause disabling the whole network if a backup up server doesn't take over immediately.

\subsection{Client-server}
The Client-server architecture is the most classical model, in which there is always a request from one side and a reply from the other. Determining device's role is challenging because sometimes we encounter devices that can be considered as both. CaSP used this model to separate processing ad sensing from each other? \cite{6512846}. This architecture can be classified into two models based on the functionality of clients.

Thin client model: The server is in charge of processing applications and managing data. Here clients just provide GUI.

Thick/ Fat client model: The server is responsible for the data management. The implementation of applications and providing GUI is done by the client.

Another type of Client-server architecture is multi-tier in which main functions like presentation, application, processing, and data management is physically separated.

\begin{table*}[t]
	\centering
	\scriptsize
	\caption{Architecture and application domain of top 20 Middlewares}
	\vspace{-0.2cm}
	\begin{tabular}{|l|clcc|}
		\hline
		\textbf{Middleware} & \textbf{Architecture} & \textbf{Main Application} & \textbf{Commercialized} & \textbf{Cloud-based} \\
		\hline
		Aura &Distributed & Pervasive computing environment  &\ding{55}  &\ding{55} \\
		ABC\&S & Service-based & Car Parking automation & \ding{55} &\checkmark\\
		Capnet & Distributed, Node-based & Mobile multimedia applications& \ding{55}&\ding{55} \\
		Carriots & Service-based  & Smart city, Smart energy  & \checkmark & \checkmark \\
		CARISMA & Distributed & Mobile computing& \ding{55} &\ding{55} \\
		CHOReOS &  Service-based and component-based & Enabling large-scale, QoS-aware adaptive choreographies   & \checkmark &\checkmark  \\
		C-MOSDEN & Distributed, Component based & Resource constrained mobile devices  &\ding{55} &\checkmark \\
		COPAL & Centralized, Component based & Context provisioning  & \ding{55} &\ding{55}\\
		CoMiHoC & Centralized & Context management in MANET environment& \ding{55} &\ding{55} \\
		DropLock & Service-based & Smart Home deployment & \ding{55} &\checkmark \\
		Gaia & Distributed, Service-based & Managing ubiquitous computing habitats and living spaces & \ding{55} &\ding{55}\\
		GSN  & Distributed & Deployment and interconnection of sensor network & \ding{55} &\ding{55} \\
		Link smart & Service-based & Intelligent networked embedded systems& \checkmark &\ding{55} \\
		OpenIoT & Service-based & Smart cities and mobile crowd sensing & \ding{55} &\checkmark\\
		Rimware & Service-based  & Heart rate Monitor (HRM) and Smart lighting (SL) &  \ding{55} & \checkmark \\
		SOCAM & Service-based & Building context-aware mobile services & \ding{55} &\ding{55} \\
		ThingWorx & Service-based  & Agriculture, Smart cities and Smart buildings & \checkmark & \checkmark \\
		UPnP & Node-based & Ubiquitous mesh home networks& \ding{55}  &\ding{55} \\
		VIRTUS & Distributed & E-health caring & \ding{55} &\ding{55} \\
		Xively &  Service-based  & Home appliances connectivity and management & \checkmark & \checkmark \\
		\hline
	\end{tabular}%
	\label{tab:2}%
\end{table*}%

\section{Middleware Service Domain}
\label{service_domain}
Middleware is a solution for implementing different services in a heterogeneous environment. First, we must identify the variety of these services. There are some researchers and organizations which try to detect all potential services that middleware can present and define an approach in order to implement them. We will discuss some of the services in the following. Each of these domains may consists of multiple sub-domains.

\subsection{Information exchange and storing}
Systems that benefit from this domain (i.e. transaction systems) should provide the ability for users to pass their request to middleware for ex-changing it with other nodes or saving their information on a database without any problem. For instance, this service allows a group of operators to use a context-aware middleware to manage a smart environment. Internet of things is primarily based on ubiquitous and pervasive computing so communicating and storing data through this distributed environment require appropriate considerations.

\subsection{Data management and analytics}
This is absolutely a huge and complex domain. Users should be able to manage databases, data security, data quality, reference and master data, Metadata and other topics. By using these services there are no longer any concerns about managing and processing data, especially Big Data. Each user can implement personal policies without any concerns. Also, services like data processing and data acquisition should be presented by IoT middleware. Data analytics is another important service that the middleware provides it by extracting statistics data or visualizing obtained data and sends them to the user application.

\subsection{Object middleware}
This type of middleware which is also known as object request broker allows applications to transmit objects and demand services via an object oriented system. In summary, these middlewares manage and control communication among objects. Remote Procedure Call (RPC) in combination with these middleware makes Distributed Object Middleware (DOM). This added feature calls objects and procedures on remote systems and can implement synchronous or asynchronous interactions among objects, applications, and systems.

\subsection{Communication}
Actually, this domain is a baseline for many other domains. A communication middleware provides a framework or environment which enables two applications to negotiate and exchange data in a distributed system. Communication middleware provides an abstraction of the network protocol for software application and reduces complexity of designing low-level communication mechanisms. DDSS is an example for communication middleware \cite{4494018}.

In addition to the above mentioned domains and by considering system services and operations, there are some other special-purpose applications which need various features of middleware domains to perform their task properly. For example WebCrawler \cite{Wu:2012:WCM:2389936.2389949} and GRank \cite{Taha2013} are two search engines middleware that uses a combination of aforementioned domains capabilities. 

\section{Comparison of Sample Middlewares}
\label{middlewares}
In this section, we describe in detail top 20 middlewares (mentioned in Table \ref{tab:2}) and provide a comparison between their main functionalities and applications from different aspects (Table \ref{tab:3}). These platforms are designed for specific applications but there are middlewares such as LinkSmart that offer a wide range of services for different needs. These middlewares have been selected for various reasons including being state-of-the-art, covering all architecture styles, working on different application domains, being commercialized and using different infrastructure were some of our selection criteria. 

\begin{table*}[t]
	\centering
	\scriptsize
	\caption{Summery of top 20 Middlewares Comparison (NS stands for Not-Specified and S\&P for Security \& Privacy).}
	\vspace{-0.2cm}
	\scalebox{0.93}{
		\begin{tabular}{|l|cccccccc|}
			\hline
			\textbf{Middleware}  & \textbf{Event Detection} & \textbf{Service Discovery} & \textbf{Adaptability} & \textbf{Platform Portability} & \textbf{Interoperability} & \textbf{Context Awareness} & \textbf{S\&P} & \textbf{Real-time}  \\
			\hline
			Aura  & \ding{51} & \checkmark  & \ding{51} & \ding{51} & \ding{55} & \ding{51} & \ding{55} & \ding{55}\\
			ABC\&S & \checkmark  & NS &\checkmark & \checkmark & NS & \checkmark &\checkmark  &\checkmark  \\
			Capnet & \ding{51} & \checkmark & \ding{51} &\ding{51} & \ding{51} & \ding{51} & NS & NS \\
			Carriots  & NS &\checkmark & \ding{51}& \ding{51} & \ding{51} & NS & \checkmark &\checkmark \\
			CARISMA  & \ding{55} &\ding{55} & \ding{51} & \ding{51} & \ding{51} & \ding{51} & \ding{55} &\ding{55} \\
			CHOReOS  & \checkmark  &\checkmark   & \checkmark  & \checkmark  &\checkmark  & \checkmark & \checkmark & \checkmark\\
			C-MOSDEN  & \checkmark  &\checkmark   & \checkmark  & \checkmark  &\checkmark  & \checkmark & NS & NS\\
			CoMiHoC & \ding{51} & NS & \ding{51} & \ding{51} & \ding{51} & \ding{51} & \ding{55}& \ding{55} \\
			Copal  & \ding{51} & \checkmark  & \ding{51} & \ding{51} & NS  & \ding{51} & \ding{51} & \checkmark \\
			DropLock   & NS   & NS & \checkmark  &\checkmark  & NS  & NS & \checkmark  & \ding {55} \\
			Gaia  & \ding{51} &\checkmark  & \ding{51} & \ding{51} & \ding{51} & \ding{51} & \ding{51}  & \ding{55} \\
			GSN  & \ding{51} & \checkmark  &\ding{51} & \ding{51} & \ding{55}  & \ding{55}  & \ding{51} &\checkmark \\
			Link smart & \ding{51} &\checkmark  & \ding{51} & \ding{51} & \ding{51} & \ding{51} & \ding{51} &\checkmark\\
			OpenIoT & NS &\checkmark & \checkmark & \checkmark & \checkmark & \checkmark & \checkmark & NS\\
			Rimware  & \ding{51} & \ding{55} & \ding{51} & \ding{51} & \ding{51} & \ding{55} & \checkmark & NS \\
			SOCAM  & \ding{51} & \checkmark & \ding{55} & \ding{51} & \ding{55} & \ding{51} & \ding{55} & \ding{55} \\
			ThingWorx & \checkmark & \checkmark &\checkmark & \checkmark & \checkmark & \checkmark & \checkmark &\checkmark \\
			UPnP  & \ding{51} & \checkmark  & \ding{51} & \ding{51} & NS & \ding{51} & \ding{51} & NS\\
			VIRTUS  & \ding{51} & \checkmark & \ding{51} & \ding{51} & \ding{51} & \ding{55} & \ding{51} & \checkmark \\
			Xively & \checkmark  &\checkmark   & \checkmark  & \checkmark  &\checkmark  & \checkmark & \checkmark &\checkmark\\
			\hline
		\end{tabular}%
	}
	\label{tab:3}%
\end{table*}%

\subsection{C-MOSDEN}
C-MOSDEN (Context-aware Mobile Sensor Date Engine) \cite{7397993} is a novel location and activity aware mobile sensing platform which can collect and process data without programming efforts. This is a plug-in-based middleware for mobile devices. To avoid cost in the process and storing data, save time and energy consumption (battery drainage), and reduce network traffic, this platform proposes an on-demand distributed crowd sensing platform which capture just required data based on user request and location. It consists of sensing as a service cloud platform (to supervise sensing task) and worker nodes (to perform the sensing task). Cloud middleware evaluates availability of worker nodes and send a request to selected ones, and also can impose specific condition on the data acquisition or transfer such as sense when certain activity occurs. It includes three main modules of context-aware, activity-aware and location-aware that work with GSN (Global Sensor Network) as cloud-based companion platform in an integrated system. Context-aware data streaming engine called mobile sensor data engine is based on previous MOSDEN platform. It is client side tools install on any device.

C-MOSDEN also have context-aware functionality and supports push/pull data streaming. Activity aware module can recognize 6 activities of I) moving in a vehicle II) cycling III) walking IV) running V) still (not moving) VI) tilting (falling) and a combination of them. Location aware module recognizes when a device enter or exit from an area which is defined by longitude, latitude, and radius. GSN aims at providing flexible middleware to address sensor data integration and distributed query processing. It is based on four basics principles: A) simplicity B) adaptability C) scalability D) lightweight implementation. It integrates, discovers, combines, queries, and filters through a XML based language. The key element in GSN is virtual sensors that can be any data producer or a combination of them. It can have multiple input data streams, but only have one output. First, sensor data consumer (city, researcher, doctors) submits requirement, then analysis problem and decide which sensor collect relevant data. The global task scheduler provides a strategic plan that how delegates the task to multiple worker nodes. This platform provides a high level of interoperability, scalability, usability and management of resources and costs by collecting only the relevant data.

\subsection{Xively}
Xively\footnote{https://xively.com/} platform is a capable and enterprise solution that provides a middleware for creating, managing and engaging ideal CoT. According to its functional specifications there are 3 main properties that specified as I) scalable and flexible connectivity features II) data engagement features III) management features that cover product's provisioning, monitoring, updating, and user management. Nowadays, users want operations in real-time and quickly so, Xively answered to these demands by deploying MQTT which is a messaging broker in CoT. Speed and scalability of Xively are made possible through Blueprint Template that is a flexible model for structuring devices and user's information and mapping relationships between them. Each template also provides contexts for managing real-time products and user's data fields, processing both Xively time's series and third party data. It also introduces a data logging devices that offer a full view of remote products down to specific sensors. Another feature is cross business automation that offers an opportunity to integrate products, customers, and other business systems. Xively offers a connected product management (CPM) platform which helps to capitalize the CoT. This makes modeling and connecting products easier and immediately allows them to start gathering data. With this platform, it is possible rapidly to find new features, proactively manage users and integrate obtained data with existing business systems.

\subsection{ThingWorx}
ThingWorx\footnote{https://www.thingworx.com/} is popular for its business management, big data, and analytics optimization. It is commonly used in agriculture, smart cities and smart buildings.
It supports MQTT, XMPP, CoAP and DDS for communication and TLS and AES as security protocols and includes Foundation, Utilities, Studio and kepware sections and related components.
The utilities section includes tools and Mashup Builder for defining, monitoring, management, and optimization. Mashup Builder allows the users to simply create their interactive applications, real-time dashboards, and mobile interfaces without any need for coding which reduces developer time and increases scalability.
Analytics as an analytical solution enables developers to easily find the real-time pattern, detect an anomaly, predictive outcomes and improve performance by using the Worx analytic server. Thing watcher automatically observes and learn normal state and alert the end user if it detects any anomaly.

By using things predictor system can predict relevant outcome. Things optimizer can improve future performance and results with automated prescription and simulation. The unique frame work allows being integrated with technologies like industrial connectivity (Kepware) for IIoT (Industrial IoT). Kepware's communications platform is a complete solution for device-to-cloud interoperability. KEPServerEX provides a single source of industrial automation data to multiple applications, allowing users to connect, manage, monitor, and control diverse automation devices and software applications through one intuitive user interface. KEPServerEX provides two options for interoperability with the ThingWorx IoT Platform. The IoT Gateway advanced plug-in features that as an agent enables real-time and read-only communication with the Platform. It allows users to model industrial things within the ThingWorx IoT Platform. The ThingWorx native client interface enables real-time and bi-directional communication with the Platform. SQUEAL (Search, Query, and Analysis) is an intelligent interactive search engine that allows a user quickly search and query and analyze through cloud data repository. It supports connectivity via 3d party device clouds, direct network connection, open APIs, and using ThingWorx edge micro server (EMS). Composer is modeling environments which make it easy to model the things, business logic, visualization, data storage, collaboration and security for application. The modeling for developer is based on making entities (things) that can produce events. To scale models and avoid repetition developer can use of thing templates (properties of things with the same nature) and things shapes (properties of things or templates).

\subsection{Carriots}
Carriot\footnote{https://www.carriots.com/} is a cloud-based IoT platform (as a PaaS) mainly designed for M2M projects like smart city and smart energy. This platform provides some modules for common M2M projects like date collection and storage, security, and device management. Seven layer architecture of the platform can provide all requirements of diverse projects and M2M applications. Carriots projects can run it in five steps of connecting the devices, collecting data, management of devices and data, building the app and running the project. It collects and stores raw streams of data from any devices by sensors through a web connectivity (gateway or embedded 3G or GPRS modems) by using MQTT protocols and send to a HTTP/ HTTPS RESTful API in XML or JSON format.

It can also integrate with other systems and pull or push data from CRM and ERP or any other APIs like Zoho and Dropbox. The platform put forward Apikeys (define privileges and visibility), HTTPS, HMAC hash solutions for security issues. Carriots include a  NoSQL big database to store data in two replication sites with high transactions. The data can be accessible through PUSH and PULL strategies of API. The powerful feature in Carriot is the capability of writing and executing code by an arbitrary code. By writing Groovy script and combining with SDK, it can execute any action in Apps and run in the cloud. Carriots device management module allows the user check status and manages configuration and firmware remotely. Carriots gives customers freedom and flexibility in matching with the wide range of hardware and some platforms such Microsoft Azure that is suitable for public cloud services. The main benefits of Carriots platform is saving development time, lower costs (in development and operation), reliability and scale up from tiny prototypes to thousands of devices (free account for up to 10 devices). It can scale up to millions of devices and customers.

\subsection{CHOReOS}
This middleware is set of software components that implement and execute large-scale web service compositions. CHOReOS tried to converge Internet of Things, Cloud Computing and the Internet of services. By using higher-level abstraction and services, this middleware tries to make scalable, complex and adaptive service structures. CHOReOS is the mixture of four main modules.
I) eXecutable Service Composition (XSC) which is responsible for coordinating the proper and required services for things.
II) eXtensible Service Access (XSA) that is used for accessing and selecting services and things.
III) eXtensible Service Discovery (XSD) which is a management framework for protocols and processes to find requested or suitable services and things.
IV) Cloud and Grid Middleware that is responsible for computational components and control implementation of choreographies. XSC have two main components 1) Composition and Estimation (C\&E) 2) Reconfiguration Management for Service Substitution. The first one is responsible for the composition of thing-based services and the second one is a composition of functional abstraction services and impact analyzers. 

XSA is the next module which includes A) XSB that backs seamless integration on heterogeneous environment B) EasyESB presents a scalable and proper environment for business services to get into large choreographies C) LSB provide services related to things and addresses IoT challenges. XSD includes I) AoSBM Discovery for resource discovery process by using service abstraction II) Things Discovery for the probabilistic method in order to implement it III) Plug-in Manager which is a flexible framework for supporting all discovery solutions. Cloud and Grid Middleware has three components 1) The Storage Service that deploy a database server on cloud and define access control policies 2) The Grid Service which provides Grid Computing as a service; this means offering high-performance computing applications enough needed resources 3) The Enactment Engine is the main section of Cloud  Grid and Middleware and is a composition of the CHOReOS middleware, business services, and coordination delegates. Cloud and Grid components export RESTful services \cite{benhamida:hal-00912882}.

\subsection{Rimware}
Rimware \cite{huo2014middleware} is a middleware proposed by Chengjia Huo et al and the main concept behind that is positioning middleware on rim that covers cloud on the inside the rim and Network of things (NoT) on the outside. The middleware propose a layer that includes gateways, clouds and components which run on the both sides. The Rimware provides scalability, interoperability, security and saving energy via gateway adapter and knowledge-based and access controller in cloud. It makes use of profile-based protocols for interoperability feature that allow devices from different vendors simply discover each other in term of their implemented profile. It uses BLE (Bluetooth 4.0 low energy) as a communication protocol that can help to lower energy consumption. First, node establish authentication and secure connection through a trusted gateway to the cloud via an adapter on gateway (when smartphone and tablets are not available), which can substitute the role of application on smart phones. Then, the cloud side establish mapping among device profiles, web APIs and cloud operations.

\subsection{DropLock}
T. L. Vinh el al. \cite{LEVINH2015234} present a middleware architecture to integrate mobile devices, sensors and cloud computing. This middleware by using cloud services provides scalability, sufficient data storage, data processing capabilities and energy saving management. The specific goal here is the convergence of Mobile Cloud Computing (MCC) and IoT. Authors presented a basic requirements framework for designing CoT middleware which includes Network management services, Data management services, System management services and Security and authentication services domains. For security concerns, before communicating with other devices, it is essential to perform the authentication process. For resource constraint devices, authentication could be performed by lightweight cryptography algorithms. Data must be encrypted before transferring on communication channels. For assurance of privacy, this middleware uses cloud-based privacy management services. The authors used a framework named DropLock to demonstrate their general service architecture on a smart city scenario.

\subsection{ABC\&S - based car parking middleware}
This middleware aimed to deploy Always Best Connected and best Served (ABC\&S) paradigm to design a worthy cloud-based car parking middleware for Internet of Things. Authors used smart mobile devices, low-powered processing chips, cloud computing and future network and communication environments (i.e. UCWW) as main enabling technologies to form their idea. This article proposed a three-layer architecture which consists of the sensor layer, communication layer, and application layer. Sensor layer consists of multiple sensor technologies such as RFID and CCTV. Communication layer is a composition of different wireless technologies (i.e. WIFI, ZigBee, VANET, and WSN). The application layer is responsible for providing some cloud-based services for actions like finding best parking lot available, vehicle license plate patrolling, car tracking and etc. This layer is a three-tier model which includes of Cloud tier, Web server tier, and Mobile app tier. In the cloud tier, web applications arranging data into an open-source software framework like Hadoop for storing them and running applications on distributed hardware. In web server tier it uses OSGi (a java framework) to dynamically publish, discover and bind services to Bundles. Mobile app tier makes mobile devices to access web applications to gain best services. In cloud tier, multiple servers may send recommendations about best available parking lots for users therefore, a cloud-based middleware developed with three clusters, A) Kafka a messaging platform for load balancing, B) Storm that consumes topics produced by Message Queue and process data (i.e. filtering, mining, clustering, etc), C) HDFS that contains useful dataset which web applications can access to them in real-time by put, scan, add, get operations \cite{s141222372}.

\subsection{OpenIoT}
OpenIoT (open source Internet of Things) \cite{Soldatos2015} is an open source IoT platform with semantic interoperability in the Cloud. This platform as a standard-based model for physical and virtual sensors applies Semantic Sensor Networks (SSN) ontology for semantic unification of IoT systems. Discovering and collecting data from mobile sensors is achievable through a publish/ subscribe middleware called Cloud-base Publish/ Subscribe (CUPUS). The CUPUS interact to cloud database via X-GSN (eXtended Global Sensor Network).
The architecture of OpenIoT includes seven elements including, a sensor middleware, cloud data storage, Scheduler, Service Delivery and Utility manager (SD\&UM), Request definition, Request presentation and Configuration and monitoring. For security and privacy issues, the platform adopts a flexible and generic approach for authentication, authorization and User management through its specific privacy \& security module and CAS (Central Authentication Service) service. To ease the development of applications (zero-programming) and manage IoT applications, it offers an integrated development environment (IDE) comprises a range of visual tools. OpenIoT is recommended for application in areas where semantic interoperability is required. The ability to handle mobile sensors and providing QoS parameters makes it a suitable choice for smart cities and mobile crowd sensing.

\subsection{Aura}
This middleware is suitable for pervasive computing. It applies two broad concepts. First is proactively which means system's layers are able to answer the request from a higher level. Secondly is self-tuning that layers must adjust their performance and resource usage according to demands made on them. The personal aura acts as a proxy and by mobility, it provides adaptation to the new environment. In Aura, the context observer helps to do context managing. One of the main challenges here is seamless integration rather than building blocks of pervasive computing. In order to achieve maximum capabilities of a resource-limited mobile client and thus improve user experiences, Aura uses cyber foraging. Surrogates are nearby computing servers or data-staging servers which provide this amplification. The reliability and quality of connections between Surrogates and nodes are important \cite{1300425}.

\subsection{Capnet}
Capnet
is mainly developed for mobile multi-media applications. This middleware is context-aware and with a common interface for context sensors, it makes it easier to embed a new sensor in the system. Capnet tries to address requirements such as mobility of the software components, multimedia applications, and adaptation. Context-aware pervasive networking focus on adapting service application according to user's environmental position and his personal preference. Service discovery and component management also provide the required level of transparency for applications making them able to perform operations in the mobile environment. This middleware is capable of asynchronous messaging, context management, service discovery, resource management and etc. Capnet can perform a proper task according to component mobility; it can decide whether to run a component on the client side (mobile device) or server side (fixed PC) depending on the resource requirements of the starting application or component \cite{Davidyuk:2004:CMM:1052380.1052410}.

\subsection{CARISMA}
CARISMA use the principle of reflection in order to achieve a dynamic behavior to context changes. It provides primitives for developers which describe how context changes should be handled using policies. For example, for message exchanging in this middleware, there are four policies: charMsg, which sent one character each time, plain Msg, to exchange messages explicitly, compressed Msg to exchange compressed messages, and encrypted Msg to deliver encrypted messages. It also enhances the construction of adaptive and context-aware mobile applications. Authors believe middleware platforms must support both deployment time configurability and run-time re-configurability. CARISMA presents an approach to provides applications, the highest economic quality of service as possible which they can offer to consumers \cite{1237173}.

\subsection{LinkSmart}
LinkSmart is an ambient intelligent (AmI) middleware that acts and reacts based-on object presence. In general, this technology has five main characteristics as follows: Context-aware, Embedded, Adaptive, Personalized, and Anticipatory. LinkSmart uses a lower-level Data Acquisition Component to collect accurate data from Context Providers. Data Acquisition Component performs two main protocol, push and pull. Push handles data that needs to be sent and pull retrieved data from sources. This middleware aims to support both distributed and centralized architectures. It will provide reflective properties, security and trust and model-driven development of applications. LinkSmart also allows for secure, trustworthy, and fault-tolerant applications with the use of distributed security and social trust components. The main domains which end-users can use are facility management, smart homes, and health care \cite{5600342}.

\subsection{GSN}
Global Sensor Networks as its name indicate, make a rapid and simple deployment of different sensor network technologies. The four main design goals are simplicity, adaptively, scalability and lightweight implementation. It helps to make a flexible integration and sensor networks discovery. GSN will provide dynamic adaption of the system configuration during operation. It uses a container-based approach which allows different sensors easily to be identified and most of the system complexities hide in that container. These containers communicate to each other in a peer-to-peer style.  The main abstraction in GSN is the virtual sensor. Virtual sensor can be any device which produces data, i.e. a real sensor, wireless camera, PC, cell phone and etc.VSM is responsible for multiple tasks such as accessing virtual sensors, controlling the delivery of sensor data, and providing the necessary administrative infrastructure. VSM has two components, LCM, and ISM. LCM manages the usage and communication of resources and virtual sensors. ISM is responsible for managing and quality streams so it ensures QoS of streams \cite{Aberer:2006:MFF:1182635.1164243}.

\subsection{COPAL}
Copal is an adaptive context provisioning approach. Context provisioning refers to the approach of collecting, transferring and processing context in order to setup context-awareness of ubiquitous services. Copal is designed to provide a runtime middleware. This will lead to loose-coupling between context and its processing which is for integrating new context sources, creating new information models and supporting different information processing requirements of context-aware services. This middleware has two key concepts. The first one is Publishers which are device services that indicate all sensors and devices in the environment. Complexity and heterogeneity of devices and communications will be solved by using wrappers. The second one is Listeners that refer to Context-aware services which COPAL must notify when corresponding inquiries are met. Each service may have numerous listeners ?\cite{5645051}.

\subsection{Gaia}
Gaia is a distributed middleware architecture that coordinates software services and heterogeneous networked physical devices. It provides a template to develop user-centric, multi-device, resource-aware, context sensitive, and mobile applications. Gaia main components are the kernel, the application framework, and the applications. Kernel component management core and application framework let users build, execute, or adapt existing applications to active spaces. Gaia offers 5 basic services: Presence service, Event manager, Context service, Space repository and Context File System.  It focuses on the interaction among users and active spaces. Active space is a programmable pervasive computing environment which boosts mobile users? ability to interact and configure multiple devices and services simultaneously. Active spaces are usually determined for specific tasks. Applications can use task context to detect meaningful information from incoherent data \cite{1158281}.

\subsection{UPnP}
The main goal here is designing a middleware for mesh home network. The wireless mesh home network consists of two kinds of devices: Mesh-Controller and Mesh client. MeshController sets a ubiquitous heterogeneous network within the house, which can be interconnected with IP net-work and limits the access to the wireless mesh home network and provides secure communications by deploying various mechanisms. A home getaway connects these Mesh Controllers together and provides an easy and centralized users' management. UPnP offers peer-to-peer ubiquitous network connectivity between various devices. Service discovery and context-awareness are two critical capabilities of UPnP. Typical UPnP protocol doesn't support sufficient context-awareness. When the mesh client moves from one point to another, the user context must be reconfigured in order to keep context-aware service discovery. The authors introduce an approach to utilize UPnP in order to make a middleware which considers user profile and context in an intermediate node. Context aggregators provides consistent and uniform contextual information to the middleware Context Management Module is responsible for creating user's related context information? \cite{5449966}.

\subsection{CoMiHoC}
CoMiHoC is suitable for context management and situation reasoning in MANET environment. Context-awareness has to establish in a peer-to-peer manner in order to work in MANET environment. CoMiHoC builds location models and estimates the connection of contexts used for situation reasoning. This middleware defines a concept of context relevancy that is specified as the rating to which particular context information is suitable to the current condition of a context-aware application. CoMiHoC architecture framework is divided into three different groups of component: Context provisioner, Request manager, and Situation reasoner. Context provisioner group provision relevant contexts to the submitted situation spaces and also manage context buffer which replaces less relevant context with more relevant one if they are available in the current position. Request manager is responsible for incoming context queries and will reply to them if they exist in the buffer. Situation reasoner consists of previous component group and manages a set of situation spaces. The authors consider challenges such as temporal relevancy, context uncertainty, distributed control and fault-tolerant which needs to be addressed in managing context in MANET environment \cite{5474762}.

\subsection{SOCAM}
In SOCAM, context is presented as ontology markup language. The benefit of this approach is that the context information can be shared among devices or entities in a pervasive computing domain and context reasoning becomes possible. By consideration of mobile device limitations and reducing the burden of context processing from them, the context model is designed in a two-level hierarchy. Pervasive computing domain divides to multiple sub-domains and each domain consist of individual low-level ontology. For linking up all sub-domains, there is also a generalized ontology.
Each component in this middleware represents as an autonomous service. Its main goal is to work on mobile and pervasive computing system. It will nearly respond to every context-based need. Service-oriented context-aware middleware is capable of doing tasks like acquiring, discovering, interpreting, and accessing various contexts and interoperability between various Context-aware systems \cite{1391402}.

\subsection{VIRTUS}
VIRTUS middleware is aiming to benefit IoT paradigm for e-health solutions. E-health features include tracking, identification, authentication, data collection and sensing. The main goals of this system are collecting patient data, monitoring his/her health status and act as a prevention. Instead of deploying SOA, it uses an Instant Messaging Protocol (XMPP)-based middleware which tries to provide a real-time, safe and trustworthy communication channel among heterogeneous devices. VIRTUS is a publish/subscribe middleware, with having XMPP on the side, it will allow the users of the system to be acknowledged if a message is arrived at the destination or not. In addition, it can inform entities if the destination user is offline. VIRTUS can support three types of devices: resource rich devices (like PC), resource-poor devices (like tablets) and simple devices (like sensors) \cite{6296229}?.

%

\section{Challenges and Issues}
\label{challenges}
This section discuss the main challenges that exist on using middlewares technologies in CoT applications. All the mentioned challenges are interesting problems and open issues that need further investigation and have great potential to be considered as future directions in this study.

\subsection{Near Real-Time Prioritizing}
CoT is a massive network of intelligent objects which all of them provide or receive resources to fulfill their tasks. There are situations that resource/service allocation should be prioritized in real-time. Most of the CoT-based middlewares are not capable prioritizing unforeseen tasks in realtime. For example in a CoT-based hospital, middleware should be able to prioritize the patients' care based on their sickness in real-time manner and retrieve patient data that is in worse condition to be notified on doctors' platform. Therefore middlewares need to apply mechanisms to measure unforeseen situation in real-time and provide cloud services according to the importance of the requests.

\subsection{Proper Resource Discovery Implementation}
CoT is a heterogeneous environment with the most dynamic topology changes. There is no guarantee for continuity of services/resources. There are so many factors that make Resource Discovery challenging in this type of networks (i.e. Nodes add to the network or leave it suddenly, the burstiness nature of network, Asynchronous interface protocols, Mobility, etc). Recent researches tried to answer some of the requirements but they were not comprehensive methods or solutions. For this purpose, moving toward more capable resource/service discovery mechanisms which provide expected performance is strictly necessary. At the moment, the most widely used discovery protocols are XMPP, UPnP and WSDiscovery.

\subsection{Users Security and Privacy Enhancement}
Implementing trustable security and privacy methods is one of the major challenges in pervasive networks. Some middlewares use their local security capabilities to evaluate the system safety but, when we use cloud services it would arise some concerns. All CoT-based middlewares that we mentioned rely on security measures that cloud provider have guaranteed. A security management platform on user-side that provides the monitoring and controlling measures on cloud-side operations would be essential. Privacy is equally important because CoT is a composition of ubiquitous devices and they are exposed to malicious devices. Common authentication and identification methods are not efficient due to frequent context changes, devices resource limitations, spontaneous device acquisition and etc.

\subsection{Supporting Various Interface Protocol for E-healtcare domain}
E-health can utilize CoT to provide various services such as patients monitoring. However, designing a suitable middleware on this sensitive domain demands more efforts. Several CoT sensors, actuators, hubs or in general Things may be implanted into patients? body or carrying by them as a wearable device. There has been inadequate attention on communication signals harmful effects on humans body. Developers need to put additional efforts to design a middleware which can support the maximum flexibility in deploying various interface protocols. For example, authors in \cite{7032050} investigated millimeter-waves frequencies side effects and threats on humans body.

\begin{table*}[t]
	\centering
	\small
	\caption{CoT-based Middleware vs. non CoT-based Middleware (``Tie" shows the feature is equally available in both CoT and non-CoT scenario and there is no competitive advantage in each-one)}
	\vspace{-0.2cm}
	\scalebox {1}
	{
		\begin{tabular}{|p{1.7cm}|cc|p{10.7cm}|}
			\hline
			\textbf{Functions} &\textbf{CoT} & \textbf{non-CoT} & \textbf{Description} \\
			\hline
			Adaptability &  \multicolumn{2}{c|}{Tie}  & In CoT-based Middleware, dynamicity for adaptability is high but non CoT-based Middleware is quicker for critical situations. \\
			\hline
			Connectivity Convergence & \checkmark & - & Cloud can provide more variety in interface protocol. \\
			\hline
			Context Management & \checkmark  & - & In Context-aware platforms, Context Management needs continuous processing that residing it in the cloud allows Middleware to save processing power.\\
			\hline
			Energy Efficiency & \checkmark & - & It could be a tie because CoT-based Middleware consumes more energy for exchanging messages but non CoT-based Middleware use more processing power and in general the later one is more considerable. \\
			\hline
			Flexibility & - & \checkmark & In non CoT-based Middleware, developers can provide services as they want and it is third-party platform- independent. \\
			\hline
			Interoperability & \checkmark & - & Cloud inherently is interoperable therefore can provide better range of heterogeneous devices. \\
			\hline
			Maintainability & -     &\checkmark      & In crisis time, it is essential for systems to perform appropriate operations in a fraction of time. non CoT-based Middleware is the better choice to fulfill this task. \\
			\hline
			Platform Portability &\checkmark   & -     & IoT is the composition of heterogeneous devices. Encountering unknown platforms is possible in non CoT-based Middleware due to weak forecasting or resource limitation to cover all platforms. Cloud can address these type of issues.\\
			\hline
			Quality of Service & -  & \checkmark   & Providing QoS in non CoT-based Middleware is more reasonable because there is more control on infrastructure.\\
			\hline
			Real-Time Tasks & -  &\checkmark   & non CoT-based Middleware performs this feature better because there is less latency on performing tasks. \\
			\hline
			Resource Discovery & - & \checkmark   & Resources change at any moment. We can refer situations that resources lifetime in network is less than cloud-based Resource Discovery process time.\\
			\hline
			Reusability & \checkmark   & -  & SoA is the main architecture style in CoT-based Middleware. This Architecture is known for reusability. CoT uses everything as a service scheme that is a heaven for reusability feature. \\
			\hline
			Security and Privacy & \multicolumn{2}{c|}{Tie} & Cloud services implement security much safer (i.e. the possibility of DDOS attack reduces significantly). In contrary, Non CoT-based Middleware is reasonable for privacy. So a tie is a fair result.\\
			\hline
			Transparency & \checkmark  & - & Transparency and abstraction is more sensible in Cloud environment. \\
			\hline
			Trustworthiness Management & \checkmark   & - & This feature is crucial in Social Internet of Things. So we expect an enormous environment and a high volume of data exchanged. Therefore using cloud is much better. \\
			\hline
	\end{tabular}}%
	\label{tab:4}%
\end{table*}%



\subsection{A light weight CoT-based Middleware for Ultra resources constrained  devices}
Internet of Nano Things (IoNT) is another IoT paradigm that is growing fast. It can be forecasted that this technology shapes a considerable proportion of IoT in near future. There are several
reasons that IoNT has attracted the attention of researchers and investors but, the main reasons are cost efficiency and high deployment density which increase system accuracy. As we can expect, for implementing this paradigm, it needs ultra resource-constrained devices that have limitations in processing capability, stored energy, storage capacity, communication range and etc. So designing a light-weight dedicated CoT-based middleware can resolve many implementations issues.

\subsection{Finding optimal place for analysis data by Fog Computing}
\begin{figure*}[t]
	\centering
	\includegraphics[width=0.8\textwidth, height=0.6\textwidth]{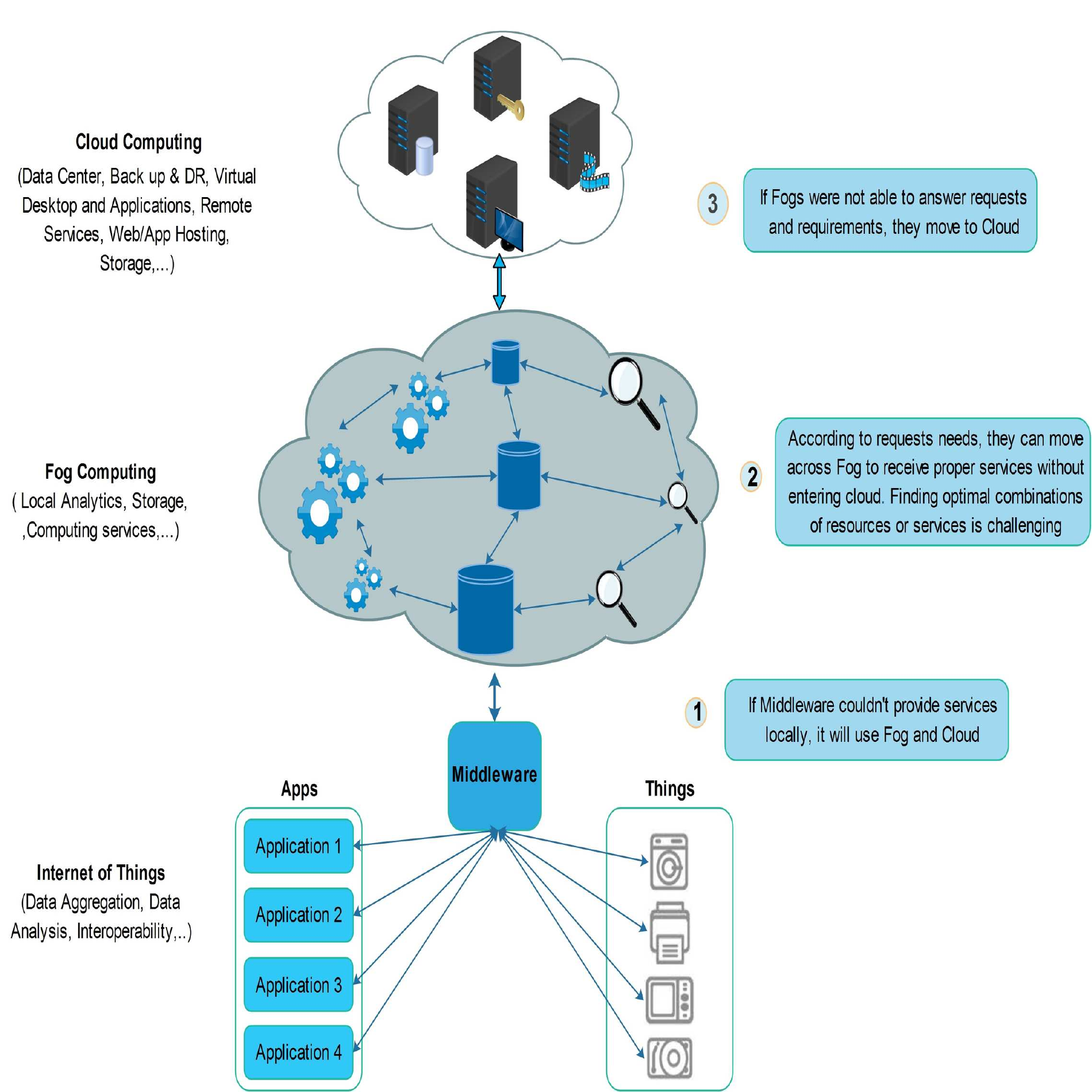}
	\vspace{-0.1cm}
	\caption{Implementing CoT-based Middleware by using Fog Computing}
	\label{fig5:Fog}
	\vspace{-0.3cm}
\end{figure*}

Fog-Computing \cite{CPE:CPE3485} has potential to fulfill some of the requirements of CoT such as lower latency, mobility support, and location-awareness. In this regard, authors in \cite{7466912} reviewed Fog Computing in CoT and underlined Fog as a middleware in Cloud Computing. Fog provides computation, storage, and networking services and stands between end nodes of IoT and traditional Clouds. One of the key functions is finding optimal place for data analysis by Fog Computing. As shown in Figure \ref{fig5:Fog}, the Fog layer take place between IoT and Cloud and tries to be close as possible to underlying system for better real-time performance. On the other hand, the variation of requested task from middleware is significant. For instance, in real-time traffic classification, middleware may receive elastic, semi-elastic, soft real-time or hard real-time requests. Determining to what extent requests can move into Fog or Cloud is challenging. Passing them to the nearest or furthest distance is not always the most effective option. As the variation of services in Fog is considerable, therefore it is necessary to perform some accurate calculations and assessments to determine optimal selection or combination of resources for requested operations. Except for real-time classification, there are many other factors which need Fog to find the most efficient place for processing them. The recent works were not really successful to implement this feature. Realization of this feature significantly impacts cloud computing position in industrial market.

\subsection{Providing Quality of Service}
In CoT, Providing SLA and guaranteeing QoS is quite challenging. Here, dynamicity of the network is really high and predicting system behavior is a big problem. Furthermore, Cloud limitations even make the situation worse. As the main requirements of QoS, we can name Bandwidth, Delay, Jitter and Packet Drop. The nature of the Cloud prevents to address those requirements. In order to implement a desired level of QoS, Fog Computing can help by filling the gaps with better time service (Table \ref{tab:5}). However, these are just basic requirements of providing QoS and researchers need to design proper mechanisms to deploy them efficiently.

\begin{table}[t]
	\centering
	\small
	\caption{Cloud \& Fog Computing QoS Requirements}
	\vspace{-0.3cm}
		\scalebox{0.94}{
	\begin{tabular}{|l|cc|}
		\hline
		\textbf{Requirements} & \textbf{Cloud} & \textbf{Fog} \\
		\hline
		\textbf{Latency} & High  & Low \\
		\textbf{Jitter} & High  & Very Low \\
		\textbf{Distance} & Through Internet & Limited Hop Count \\
		\textbf{Hard Real-time} & No or Limited & Yes \\
		\textbf{Mobility} & Limited & Supported \\
		\textbf{Bandwidth} & More Demand & Less Demand \\
		\hline
	\end{tabular}%
	}
	\label{tab:5}%
\end{table}%

\subsection{Standardization needs}
Based on the conducted literature studies, we found out that one of the main and obvious problems is the lack of standardization in designing approaches. As a suggestion for future work, a proposal of layered-based model can be helpful and provides significant stability. although, by considering heterogeneity nature of Internet of Things it may seem quite challenging and add more complexity to the already complex scheme.

%


\emph{In a nutshell and as concluding remarks of this study, we present a head to head table (Table \ref{tab:4}) which compares different features of CoT-based and non CoT-based Middleware. As we found, CoT-based Middleware are suitable in several usecases including social, smart home, smart city, smart agriculture, smart animal farming, smart grids and smart retail applications which produce massive data and don?t need hard real-time processing.
	In contrary, non CoT-based middlewares can be more utilisable in industrial control, E-healthcare and smart logistics domains which need hard real-time processing. }

\section{Conclusion}
\label{Conclusion}
Cloud of Things (CoT) platforms will play a key role in near future for enabling service provision in large Internet of Things environments. Toward that goal, understanding the role and necessity of middlewares in CoT is an essential.
This study focuses on middleware technologies in CoT in three main directions. Firstly it presents the major features and characteristics of middlewares in CoT domains and compare them from different architectural design possibilities.
Next, several middlewares are introduces and studied based on their architectures, service domain, application.
Lastly a list of open challenges and issues are presented that are interesting problems to be tackled as future direction of this study.

\end{document}